\documentclass{PoS}

\title{Search for new physics with the SHiP experiment at \textsc{cern}}

\ShortTitle{Search for new physics with the SHiP experiment at \textsc{cern}}

\author{\speaker{Oliver Lantwin}\\
       Imperial College London, London, United Kingdom\\
       E-mail: \email{oliver.lantwin@cern.ch}}

       \abstract{SHiP is a new general purpose fixed target experiment at the \textsc{cern} 
         \textsc{sps} designed to complement \textsc{lhc} experiments in the search
        for new physics.  In its initial phase, the \(400\)~GeV proton beam
        extracted from the \textsc{sps} will be dumped on a heavy target with
        the aim of integrating \(2\times10^{20}\)~pot in 5 years. Shielded by
        an active muon shield, a dedicated detector, based on a long decay
        volume followed by a spectrometer and particle identification
        detectors, will allow probing a variety of models with light long-lived
        exotic particles with masses below \(\mathcal{O}(10)\;
        \mathrm{GeV}/{c^2}\). The main focus will be the physics of the
        so-called Hidden Portals, i.\,e.\ search for Dark Photons, Light
        scalars and pseudo-scalars, and Heavy Neutral Leptons. The sensitivity
        to Heavy Neutral Leptons will allow for the first time to probe, in the
        mass range above the kaon mass, a coupling range for which Baryogenesis
        and active neutrino masses could also be explained.  A dedicated
        emulsion-based detector will allow detection of light dark matter in an
        unexplored parameter range.}

\FullConference{The European Physical Society Conference on High Energy Physics\\
         5--12 July\\
         Venice, Italy}

\usepackage[utf8]{inputenc}
\usepackage{booktabs}
\usepackage{cite}
\usepackage{xspace}
\usepackage{subcaption}
\usepackage{etex}
\usepackage[alsoload=hep]{siunitx}
\usepackage{amsmath, amssymb}
\usepackage{slashed}
\usepackage[british]{babel}
\usepackage{mathtools}
\usepackage{acro}
\usepackage{lineno}
\usepackage{microtype}

\DeclareAcronym{nuMSM}{
  short = \ensuremath{\nu}MSM\xspace ,
  long  = neutrino minimal standard model ,
  class = abbrev
}
\DeclareAcronym{pid}{
  short = \textsc{pid}\xspace ,
  long  = particle identification ,
  class = abbrev
}
\DeclareAcronym{ship}{
  short = SHiP ,
  long  = Search for Hidden Particles ,
  class = abbrev
}
\DeclareAcronym{hnl}{
  short = \textsc{hnl}\xspace ,
  long  = heavy neutral lepton ,
  class = abbrev
}
\DeclareAcronym{sm}{
  short = \textsc{sm}\xspace ,
  long  = standard model ,
  class = abbrev
}

\DeclareAcronym{bsm}{
  short = \textsc{bsm}\xspace ,
  long  = beyond standard model ,
  class = abbrev
}

\DeclareAcronym{sps}{
  short = \textsc{sps}\xspace ,
  long  = super proton synchrotron ,
  class = abbrev
}
\DeclareAcronym{cern}{
  short = \textsc{cern}\xspace ,
  long  = Conseil Européen pour la Recherche Nuclèaire ,
  class = abbrev
}
\DeclareAcronym{tp}{
  short = \textsc{tp}\xspace ,
  long  = technical proposal ,
  class = abbrev
}
\DeclareAcronym{pp}{
  short = \textsc{pp}\xspace ,
  long  = physics proposal ,
  class = abbrev
}

\begin{document}

\section{Introduction}\label{introduction}

\acuse{cern}

Particle physics is faced with a puzzle. Even though there is experimental
evidence for new physics beyond the \ac{sm}, with the exception of
neutrino oscillations and indications from cosmology, so far all laboratory
experiments have been incredibly consistent with the predictions of the
standard model. This leaves us clueless as to where the \ac{sm} breaks down
and \ac{bsm} physics takes over. And so far none of the predictions of popular
\ac{bsm} models have been confirmed.

There are two possibilities for why we did not see \ac{bsm} physics yet:
The \textsc{lhc} might not be powerful enough to explore the scale of new physics,
which could be just around the corner, but could be out of reach of current and future
technology as well. Alternatively, new physics could be too weakly coupled to the \ac{sm} to be seen at
general purpose experiments.

I focus on this second option here and will explain how \ac{ship}\cite{TP} is designed to
find these elusive particles, in particularly for super-weakly coupled new
physics with \(m_\mathrm{NP} < \mathcal{O}(\SI{10}{\giga\eV})\).

If there is super weakly coupled new physics, there generally is a
\emph{portal} that mediates between the standard model and one or more
\emph{hidden} particles, i.e.~the hidden sector \textsc{(hs)}:

\[ \mathcal{L} = \mathcal{L}_\mathrm{SM} + \mathcal{L}_\mathrm{portal} +
\mathcal{L}_\mathrm{HS}, \] where \(\mathcal{L}_\mathrm{SM}\) is the \ac{sm}
Lagrangian, \(\mathcal{L}_\mathrm{HS}\) is the Lagrangian of a possibly richly
structured hidden sector, and \(\mathcal{L}_\mathrm{portal}\) are the
Lagrangian terms linking the two, i.\,e.\ those we could conceivably test for
experimentally.  If these interaction terms do exist, their mathematical form
is constrained by the fact that they, by definition, also involve \ac{sm}
fields.  The simplest possible interactions of this kind are tabulated in
table~\ref{tab:interactions}. 

\begin{table}[b]
  \centering
\begin{tabular}{l r}
  \toprule
  Portal & Interaction term \\
  \midrule
  Scalar (e.\,g.~dark scalar, dark Higgs) & \((H^\dagger H)\phi\) \\ 
  Vector (e.\,g.~dark photon) & \(\epsilon F_{\mu\nu}F^\prime_{\mu\nu}\) \\
  Fermion (e.\,g.~\acf{hnl}) & \(H^\dagger  \overline{N}
  L\) \\
  Axion-like particle \textsc{(alp)} & \(a F^{\mu\nu}\tilde{F}^{\mu\nu}\) \\
  \bottomrule
\end{tabular}
  \caption{Possible portal interactions}
\label{tab:interactions}
\end{table}

As an example to motivate the \ac{ship} design, we will consider the \acf{hnl}
of the \ac{nuMSM} --- a fermion portal.  For details on the many other models,
the reader is referred to our \ac{pp}\cite{PP}.  The
\ac{nuMSM}\cite{Asaka:2005an} is a model with a minimal number of additional
particles that can solve all of the experimental shortcomings of the \ac{sm}.
In it three right handed neutrinos \(N_i\) are added to complete the \ac{sm}:

  \begin{itemize} 
  \item A light \(N_1\) with mass
    \(\mathcal{O}(\SI{10}{\kilo\eV})\) is essentially decoupled from
    \(N_{2,3}\), making it a dark matter candidate.
  
  \item Two heavy  \(N_{2,3}\) with masses
    \(\mathcal{O}(\SI{1}{\giga\eV})\) mix with the active neutrinos,
    effectively coupling them weakly to the \ac{sm}. They are the
    \ac{hnl}. Through the mixing they set the active neutrino masses via
    the see-saw mechanism, and via leptogenesis they can also explain the
    baryon asymmetry of the universe (which is converted from an asymmetry of
    the leptons to baryons via sphaelerons). Importantly for experimental
    studies, they can be produced in heavy flavour decays, and can decay back
    to visible final states.

  \end{itemize}

With this benchmark model in mind, we can turn to how \ac{ship} is
designed to look for this and other portal models.

\section{The SHiP experiment}\label{the-ship-experiment}

\begin{figure} \centering
\includegraphics[width=15cm]{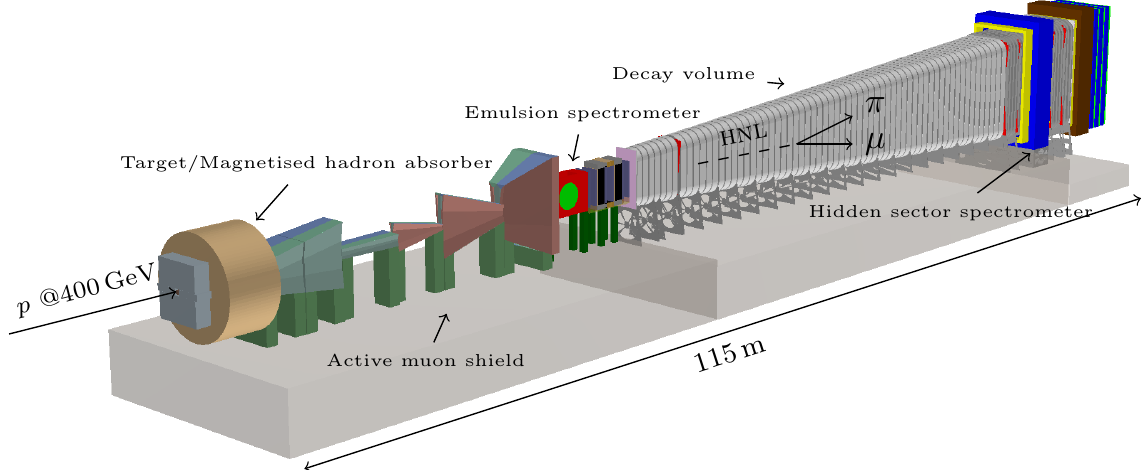}
\caption{SHiP}
\label{fig:ship}
\end{figure}

An overview of the \ac{ship} facility is shown in figure~\ref{fig:ship}.
The \ac{ship} experiment is designed to look for two types of signatures
predicted by many new physics models:
\begin{enumerate}
  \def\labelenumi{\arabic{enumi}.}
\item Via decay to visible particles in hidden sector spectrometer,
\item Via scattering off electrons or nuclei in nuclear emulsion.
\end{enumerate}

To ensure sensitivity to very weakly coupled new physics it is essential to
maximise the intensity while minimising backgrounds.  An intense proton beam from
the \ac{sps} at \SI{400}{\giga\eV} at the new beam dump facility
\textsc{(bdf)} in the North Area will supply \ac{ship} with \(2\times10^{20}\)
protons on target over 5 years. It impinges on a very dense target of
\(12\times\lambda_\mathrm{int}\), which ensures abundant heavy flavour
production while reducing neutrino production from \(\pi\) and \(K\) decays.
The yields of \(D\)-mesons and \(\tau\)-leptons over 5 years are expected to be
in excess of \(10^{18}\) and \(10^{16}\) respectively. Additionally, there will
be a high yield of photons from bremsstrahlung, \textsc{qcd} processes and
meson decay, which allows the search for e.\,g.\ dark photons.

The number of protons extracted from the \ac{sps} will be similar to that
provided for the \ac{cern} neutrinos to Gran Sasso \textsc{(cngs)} project, but
with slow instead of fast extraction of the beam. This will allow operation in
parallel with the \textsc{lhc} and other beam-lines at the \ac{sps}.

With enough intensity to possibly produce new particles, the crucial
challenge becomes rejecting backgrounds from \ac{sm} processes, \ac{ship}
aims to be a zero background experiment for the visible decay signature in the
hidden sector spectrometer.

The heavy target and the hadron absorber stop most \ac{sm} particles,
with the exception of muons and neutrinos.  Since the decay volume is under a
vacuum to prevent neutrino interactions within the fiducial volume, the muons become
the key problem. As muons lose very little momentum in material, and the
\ac{ship} detectors need to be as close to the target as possible, active
shielding is needed to deflect the muons away from the detectors.  This active
shield is comprised of a system of warm electro-magnets, which first separate the
muon charge and then deflect them to either side.  To achieve the goal of zero
background it needs to reduce the flux of muons in the detector by at least six
orders of magnitude, for the full kinematic range of muons produced, so up to
\(p\sim\SI{350}{\giga\eV}\) and \(p_T\sim\SI{8}{\giga\eV}\), in as short a
distance as possible. This makes the shield configuration critical to optimise,
and a re-optimisation using detailed simulations and Bayesian Optimisation
techniques is currently in progress. For more details on the current design,
see Ref.~\cite{muon_shield}.
For this optimisation an accurate knowledge of the muon spectrum is vital. To
improve our confidence that the muon spectrum is modelled well in simulation a
measurement of the muon spectrum for the \ac{ship} target at the \textsc{h4}
test-beam at \ac{cern}'s \textsc{sps} is planned for 2018\cite{vanHerwijnen:2267770}.

\begin{table}[b]
  \centering
\begin{tabular}{lr}
\toprule
Particle & Final states \\ 
\midrule
\ac{hnl}, neutralino & $\ell^\pm\pi^\mp$, $\ell^\pm K^\mp$, $\ell^\pm\rho^\mp$\\
Vector, scalar, axion portals; goldstino & $\ell^\pm\ell^\mp$\\
\ac{hnl}, neutralino, axino& $\ell^\pm\ell^\mp\nu_\ell$\\
Axion portal, sgoldstino & $\gamma\gamma$\\
Sgoldstino & $\pi^0\pi^0$\\
\bottomrule
\end{tabular}
  \caption{Visible final-states by hypothesised signal}
\label{tab:final}
\end{table}

To further reduce backgrounds from particles produced by muons passing through material, neutrino
interactions in the surrounding structures and cosmic rays, the decay vessel is
surrounded by background taggers, to detect any visible particles entering or
exiting the vessel.  A timing detector will further suppress combinatorial
background, while tracking is used for vertexing and impact parameter
measurement, further improving the rejection of fake signal candidates.
Finally calorimeters and a muon detector allow particle identification,
allowing study of specific final states. Some of the expected signal final
states are tabulated in table~\ref{tab:final}.  Taken together these subsystems
are designed to  redundantly reduce any possible backgrounds to negligible levels.

To search for hidden sector particles via the complementary signature of
scattering, which is particularly important for e.\,g.\ light dark matter, and to study tau
neutrinos, a detector based on nuclear emulsions is situated just downstream of
the muon shield in front of the decay volume of the hidden sector spectrometer.

\section{SHiP Sensitivity}\label{ship-sensitivity}

This section will give a brief overview of the \ac{ship} sensitivity to several
classes of models compared to other current and future experiments. Please
note, that these sensitivities are from before the current round of
re-optimisation, i.\,e.\ these sensitivity curves correspond to the \ac{tp}\cite{TP} configuration.

\begin{figure} \centering
\begin{subfigure}[t]{0.45\textwidth}
\includegraphics[height=5.5cm]{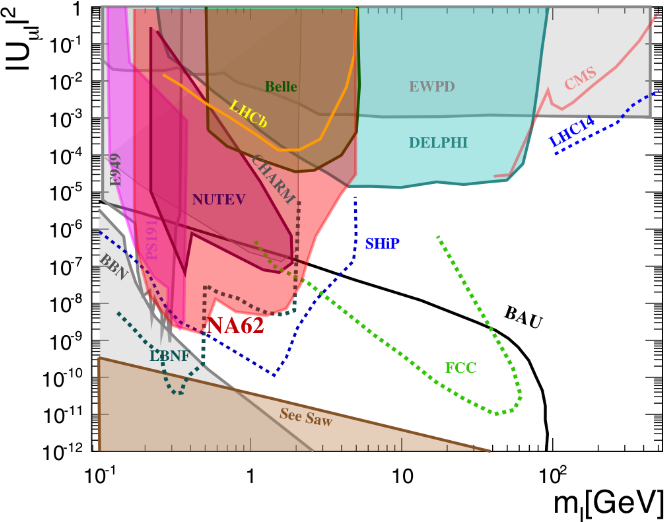}
\caption{}
\label{fig:hnl_sens}
\end{subfigure}
\begin{subfigure}[t]{0.45\textwidth}
\includegraphics[height=5.5cm]{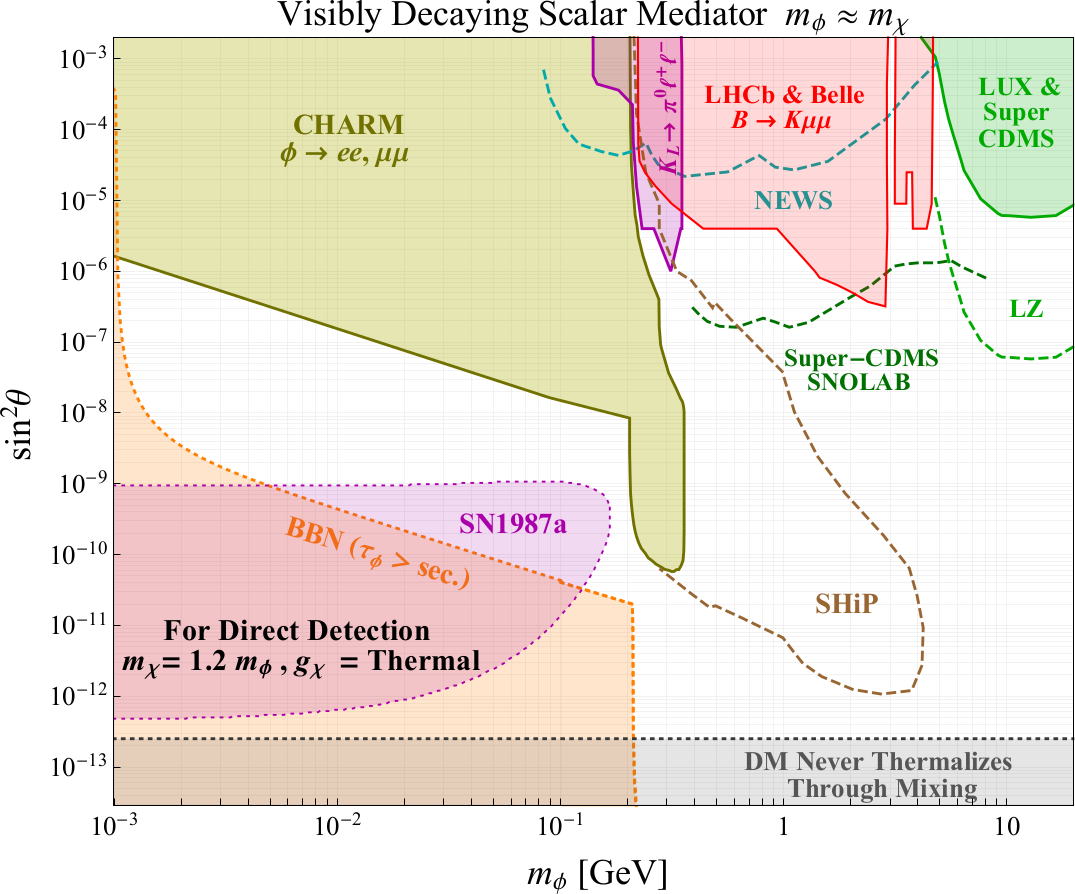}
\caption{}
\label{fig:ds_sens}
\end{subfigure}
\begin{subfigure}[t]{0.45\textwidth}
\includegraphics[height=5.5cm]{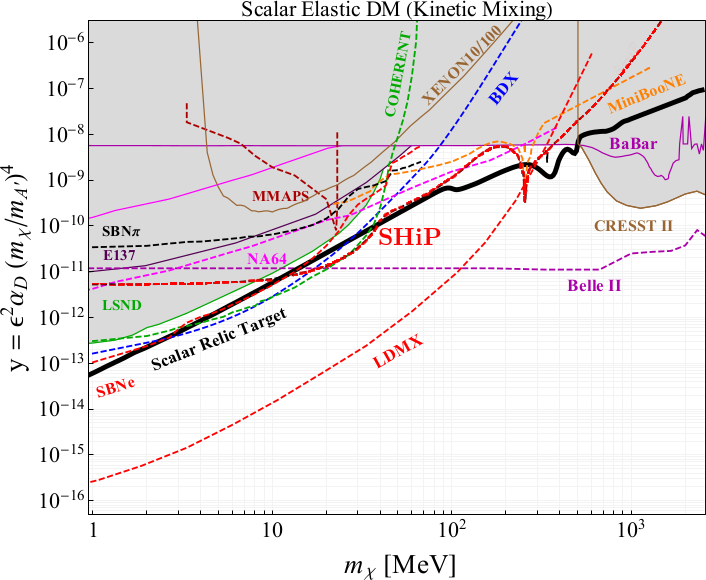}
\caption{}
\label{fig:ldm_sens}
\end{subfigure}
\caption{Sensitivity for different models at \ac{ship}. (\subref{fig:hnl_sens})
  \ac{hnl} sensitivity at \ac{ship} for \ac{nuMSM} with $U_e^2 : U_\mu^2 :
U_\tau^2 = 1 : 16 : 3.8$ and a normal neutrino mass hierarchy. Source:
Ref.~\cite{tspadaro}; (\subref{fig:ds_sens}) Dark scalar sensitivity at SHiP.
Source: Ref.~\cite{Krnjaic:2015mbs}; (\subref{fig:ldm_sens}) Light dark matter
sensitivity at SHiP for $\frac{m_{A^\prime}}{m_\chi} = 3$. Source:
Ref.~\cite{Battaglieri:2017aum}}
\end{figure}

\subsection{\ac{hnl}}

For \ac{hnl} the available parameter space is limited theoretically by
observations of the baryon asymmetry of the universe \textsc{(bau)}, the big
bang nucleosynthesis \textsc{(bbn)} and a model-independent limit for all
see-saw models. The \ac{ship} sensitivity for \ac{hnl} in this space is shown
in figure~\ref{fig:hnl_sens}.

The \ac{ship} sensitivity is best up to about \SI{3}{\giga\eV}, which is above the
charm kinematic limit, thanks to a significant contribution from \(B\)-meson
decays. In this region it is unique and complementary to the region that could
be probed at the future circular collider \textsc{(fcc)} in \(e^+e^-\) mode.

\subsection{Dark scalars}

The \ac{ship} sensitivity for dark scalars is shown in
figure~\ref{fig:ds_sens}.  Again, \ac{ship} covers a unique part of the
parameter space, complementary to other experiments. For short lifetimes
\(B\)-factories and LHCb dominate.  \(B\)-decays have a large contribution to
the sensitivity achievable at \ac{ship}. Note, that there is a gap at
\(c\tau\sim\mathcal{O}(\si{\metre})\), where the lifetime is too short for SHiP and
too long for the \(B\)-experiments, emphasising the importance for \ac{ship} to be
as close a possible to the target.

\subsection{Dark photons}

\begin{figure} \centering
\begin{subfigure}[t]{0.5\textwidth}
  \includegraphics[height=6cm]{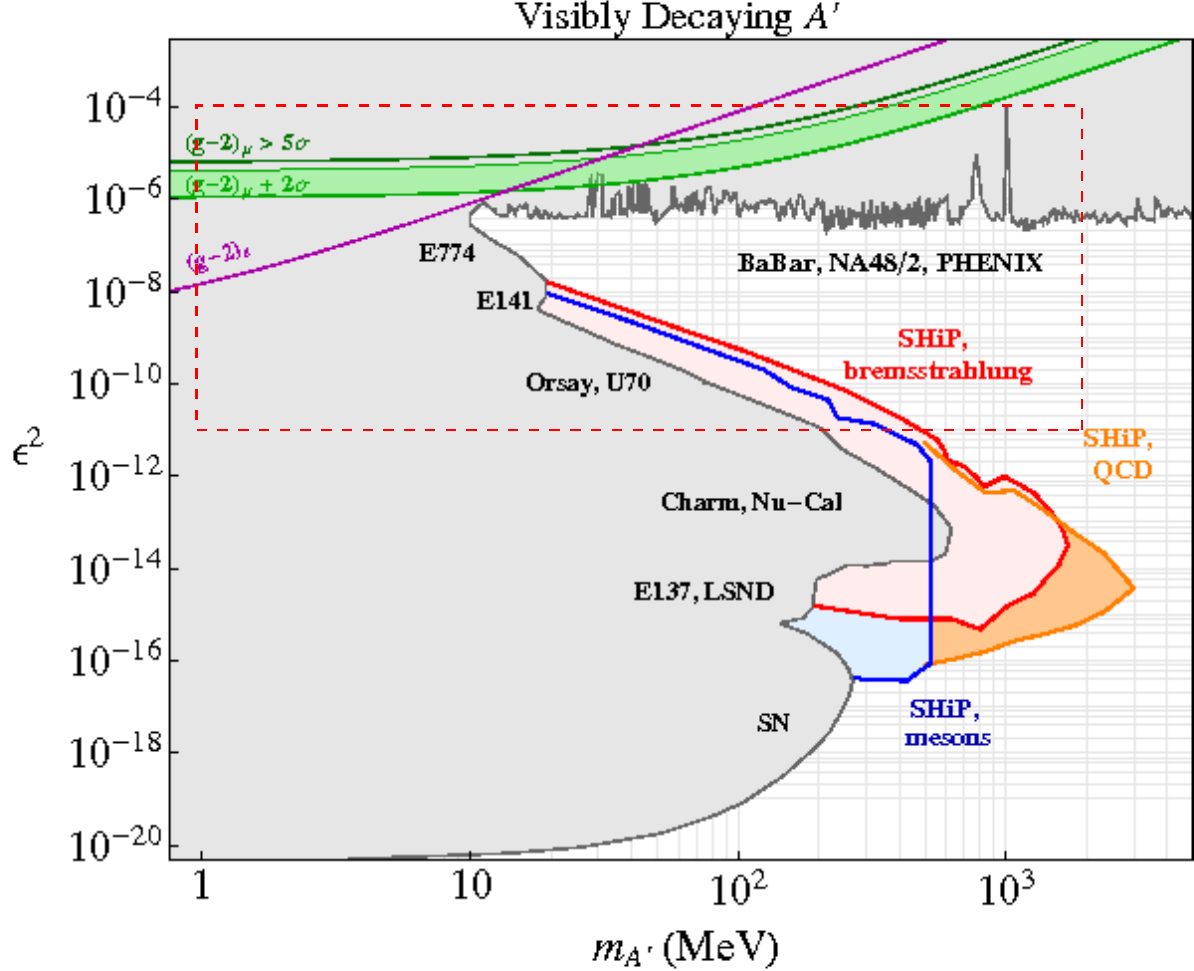}
  \caption{}
\label{fig:dp_sens_a}
\end{subfigure}
\begin{subfigure}[t]{0.4\textwidth}
  \includegraphics[height=6cm]{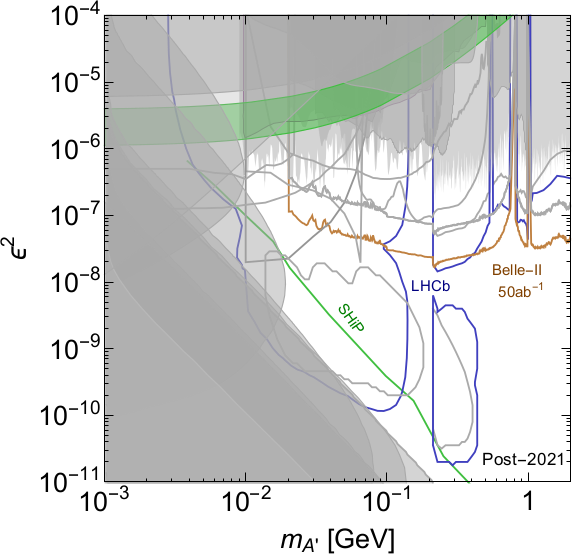}
  \caption{}
\label{fig:dp_sens_b}
\end{subfigure}
\caption{Dark photon sensitivity at SHiP. (\subref{fig:dp_sens_a}) Sensitivity to
  visible final states. Source: Ref.~\cite{PP}; (\subref{fig:dp_sens_b}) Detail of
highlighted region. Source: Ref.~\cite{Alexander:2016aln}}
\label{fig:dp_sens}
\end{figure}

The \ac{ship} sensitivity for dark photons is shown in
figure~\ref{fig:dp_sens}. This estimate is based on a yield of \(>10^{20}
\gamma\) at \ac{ship} over 5 years. The dark photons here decay to visible final
states. The \ac{ship} simulation includes production via \textsc{qcd},
bremsstrahlung and meson decays, with the respective sensitivities shown
separately in figure~\ref{fig:dp_sens_a}, but not yet via electromagnetic
showers, which are currently being implemented.

The sensitivity of \ac{ship} covers a broad region of the parameter space up to
masses of a few \si{\giga\eV} and couplings down to $10^{-16}$. The upper
boundary of the \ac{ship} sensitivity is determined by the detector's distance
to the target, as dark photons would decay before reaching the decay volume.
However, other existing and future experiments can explore this region at
higher couplings which is shown in more detail in figure~\ref{fig:dp_sens_b},
complementing the region explorable at \ac{ship}.

\subsection{Light dark matter}

For dark matter lighter than \textsc{wimp}s ``direct detection'' experiments
quickly lose sensitivity, due to the small recoil energy, which requires a very
low energy threshold of the detectors. The two common approaches to hunt for
light dark matter are via missing mass/energy searches\footnote{Missing energy
searches assume a dark photon mediator and are thus insensitive to light dark matter
produced by other mediators.} (\(\propto U^2\)) and via scattering/recoil
(\(\propto U^4\)), which are complementary. 

At \ac{ship} light dark matter is searched for indirectly via electron and
nuclear recoil in nuclear emulsion. The main backgrounds here are electron
recoils from \(\nu_e\) scattering, but differences in energy and angular
spectra can be exploited to look for an excess consistent with light dark
matter.  The \ac{ship} sensitivity for light dark matter is shown in
figure~\ref{fig:ldm_sens}. Note that the sensitivity shown here is preliminary,
as cascade production of light dark matter is not yet implemented. Furthermore,
this sensitivity projection only considers electron recoil, while nuclear
recoils are not yet included.  Even though, \ac{ship} already has the best
sensitivity for scattering, complementing \textsc{ldmx}, which searches for
light dark matter via missing energy at an electron beam.

\section{Conclusion}\label{conclusion}

There is plenty of unexplored parameter space in the dark sector new physics
could hide in. \ac{ship} is designed to be sensitive to many different final
states for both decay and scattering, allowing it to probe a vast range of
models.

Currently \ac{ship} is being re-optimised to improve the physics performance
further while respecting cost constraints. In this context the sensitivities
and backgrounds are currently being re-evaluated and updated for new
configurations. For the sensitivity updates in particular additional physics
models, production and decay channels are being added in close collaboration
with the theoretical community.

\bibliographystyle{JHEP}
\bibliography{biblio.bib}

\end{document}